\documentclass[letterpaper]{emulateapj}                %
\usepackage{apjfonts}                   
\usepackage{epsfig,graphicx,latexsym,amsmath,amssymb}                         %
\usepackage{natbib}                                                           %
\bibpunct[,]{(}{)}{;}{a}{}{,}                                                 %

\newcommand{\mbul}{M_{\bullet}}                                     %
\def\pderiv(#1/#2){\frac{\partial#1}{\partial#2}}                   %

\begin{document}

\author{Miguel Preto\altaffilmark{1}\thanks{e-mail: miguelp@ari.uni-heidelberg.de} \&
Pau Amaro-Seoane\altaffilmark{2}\thanks{e-mail: Pau.Amaro-Seoane@aei.mpg.de}}
\altaffiltext{1}
{(MP) Astronomisches Rechen-Institut, Zentrum f\"{u}r Astronomie, University of Heidelberg, D-69120 Heidelberg, Germany}
\altaffiltext{2}
{(PAS) Max Planck Intitut f\"ur Gravitationsphysik
(Albert-Einstein-Institut), D-14476 Potsdam, Germany and Institut de Ci{\`e}ncies de l'Espai, IEEC/CSIC, Campus UAB,
Torre C-5, parells,
$2^{\rm na}$ planta, ES-08193, Bellaterra, Barcelona, Spain}

\date{\today}

\title{On strong mass segregation around a massive black hole \\ Implications for lower-frequency gravitational-wave astrophysics}

\begin{abstract}
We present, for the first time, a clear $N$-body realization of the {\it strong mass segregation} solution 
for the stellar distribution around a massive black hole. We compare our $N$-body
results with those obtained by solving  the orbit-averaged Fokker-Planck (FP) 
equation in energy space. The $N$-body segregation is slightly stronger than in the
FP solution, but both confirm the {\it robustness} of the regime of strong segregation when the number 
fraction of heavy stars is a (realistically) small fraction of the total population.
In view of recent observations revealing a dearth of giant stars in the sub-parsec region of the 
Milky Way, we show that the time scales associated with cusp re-growth are not longer than
$(0.1-0.25) \times T_{rlx}(r_h)$. These time scales are shorter than a Hubble time
for black holes masses $\mbul \lesssim 4 \times 10^6 M_\odot$ and we conclude that quasi-steady,
mass segregated, stellar cusps may be common around MBHs in this mass range. Since EMRI 
rates scale as $\mbul^{-\alpha}$, with $\alpha \in [\frac{1}{4},1]$, a good fraction of these events should 
originate from strongly segregated stellar cusps. 
\end{abstract}


\section{Introduction}
The distribution of stars around a massive black hole (henceforth MBH) is a
classical problem in stellar dynamics \citep{1976ApJ...209..214B,1977ApJ...211..244L}.
The observational demonstration of the existence of nuclear stellar clusters (henceforth
NSCs)---as revealed by a clear upturn in central surface brigthness---in the
centers of galaxies makes its study ever more timely. A number of NSCs in coexistence
with a central MBH have recently been detected \citep{2009MNRAS.397.2148G} suggesting
that NSCs around MBHs, like the one in the center of the Milky Way, may be quite common.

The renewed interest in this theoretical problem is thus motivated by the observational data in NSCs and, 
in particular, the very rich and detailed data available for the stars orbiting the Galactic MBH. At the
same time,  the prospects for detection of  gravitational waves (GWs) from extreme mass ratio
inspirals (henceforth EMRIs) with future GW detectors such as the {\it Laser Interferometer Space
Antenna} (LISA) also urge us to build a solid theoretical understanding of sub-parsec structure of 
galactic nuclei. In fact, EMRI rates will depend strongly on the stellar density of compact remnants as
well as on the detailed physics within $O$($0.01$pc) of the hole, which is the region from which these 
inspiralling sources are expected to originate \citep{2005ApJ...629..362H}.

\cite{1976ApJ...209..214B} have shown, 
through a kinetic treatment that, in the case all stars are of the same mass, this quasi-steady distribution takes 
the form of power laws, $\rho(r) \sim r^{-\gamma}$, in physical space and $f(E) \sim E^p$ in energy space 
($\gamma=7/4$ and $p=\gamma-3/2=1/4$). This is the so-called {\it zero-flow 
solution} for which the net flux of stars in energy space is precisely zero. \cite{2004ApJ...613L.109P} and
\cite{2004ApJ...613.1133B} were the first to report $N$-body realizations of this solution, thereby
validating the assumptions inherent to the Fokker-Planck (FP) approximation---namely, that scattering is dominated 
by uncorrelated, $2$-body encounters  and, in particular, dense stellar cusps populated with stars of the {\it same mass} 
are robust against ejection of stars from the cusp.

The properties of stellar systems that display a range of stellar masses are only very poorly reproduced by single 
mass models.  It is well known from stellar dynamical theory that when several masses are present there 
is mass segregation---a process by which the heavy stars accumulate near the center while the lighter ones float 
outward \citep{1987degc.book.....S}. Accordingly, stars with different mass get distributed with different density 
profiles. By assuming a stellar 
population with two mass components, \cite{1977ApJ...216..883B}---hencefort BW77---generalized 
their early cusp solution and argued heuristically for a scaling relation $p_L = m_L/m_H \times p_H$ that depends 
on the star's mass ratio only. However, they obtained no general result on the inner slope of the heavy objects; nor did  
they discuss the dependence of the result on the component's number fractions. On the other hand, it was shown long ago 
by \cite{1969A26A.....2..151H} that the presence of a mass spectrum leads to an increased rate of stellar ejections 
from the core of a globular cluster, but he did not include the presence of a MBH at the center. H\'enon's work raises the 
question as to whether {\it multi-mass} stellar cusps, obtained from the solution of the FP equation, are robust against 
ejection of stars from the cusp. Ejections---due to strong encounters---are {\it a priori} excluded from the FP evolution, 
even though they could occur in a real nucleus. Furthermore, even if cusps were shown by $N$-body results to be robust 
against stellar ejections (and we show that they are in this Letter),  BW77 scaling  cannot be valid for arbitrary number 
fractions. 

Recently \cite{2009ApJ...697.1861A}---henceforth AH09---stressed this latter point and have shown via FP 
calculations that, indeed, in the limit where the number fraction of  heavy stars is realistically small, a new 
solution that they coined {\it strong mass 
segregation} obtains with density scaling as $\rho_H(r) \sim r^{-\alpha}$, where $\alpha \gtrsim 2$. They have
shown that there are two branches for the solution parametrized by 
$\Delta = \frac{N_HM_H^2}{N_LM_L^2} \cdot \frac{4}{3+M_H/M_L}$. The $weak$ branch, for $\Delta > 1$ 
corresponds to the scaling relations found
by BW77; while the $strong$ branch, for $\Delta < 1$, generalizes the BW77 solution. There is a straightforward
physical interpretation. In the limit where heavy stars are very scarce, they barely 
interact with each other and instead sink to the center due to dynamical friction against  the sea of light 
stars. Therefore, a quasi-steady state forms in which the heavy star's current is not nearly zero and thus the BW77
solution does not hold. As $\Delta$ increases, self-scattering of heavies becomes important and the resulting 
quasi-steady state forms with a nearly zero current for stars of all masses, so BW77 solution is recovered.

For all these 
reasons,  it is fundamental to verify the Bahcall-Wolf solution---as well as its Alexander-Hopman generalization---with 
$N$-body integrations. There has been a surprisingly small number of $N$-body studies of multi-mass systems around a 
MBH \citep{2004ApJ...613.1143B,2006ApJ...649...91F}, and none of them reported the occurence of strong mass 
segregation.

In this Letter we use direct $N$-body integrations to show for the first time that:  (i) strong mass segregation is a 
robust outcome of the growth of stellar cusps around a MBH when $\Delta < 1$; (ii) BW77 solution is recovered when 
$\Delta > 1$; (iii)  as a corollary, we conclude that the rate of stellar ejections from the cusp is too low to destroy the 
high density cusps around MBHs---even though ejections from the cusp {\it do} occur.  Furthermore, having 
validated the FP formalism, we proceed to use it to estimate 
the time scales for cusp re-growth starting from a wider range of models. \cite{2009arXiv0909.1318M} 
obtained, for Milky Way type nucleus, times in large excess of a Hubble time. We use a FP formalism which, in contrast 
with that of the latter author, is tailored to follow the simultaneous evolution of the cusp of different stellar masses 
without any restrictions with respect to the values of $f(E)$ or $\rho(r)$. With our FP solutions we show that, for 
$\mbul \lesssim 5 \times 10^6 M_\odot$, the times for re-growing  stellar cusps are shorter than a Hubble time. 
Our results clearly suggest that strongly segregated stellar cusps around MBHs in this mass range may be quite 
common in NSCs and should be taken into account when estimating  EMRI event rates.

\section{Models and Initial Conditions}
We have performed the $N$-body simulations with a modified version of {\it NBODY4} 
\citep{1999PASP..111.1333A,2003gnbs.book.....A} adapted to the $GRAPE-6$ special-purpose
hardware. 
The code was modified to add the capture of stars by the MBH: stars that enter
a critical radius $r_{cap}$ from the hole are captured and their mass is added to the hole. The
new position and velocity of the new massive particle are calculated by imposing that the capture 
process conserves total linear momentum.  The maximum number of particles supported
by the memory of a micro-GRAPE board is $\sim 1.2 \times 10^5$, which have been shown to be
sufficient to accurately describe the evolution of the bulk properties (densities in physical and
phase space) of the nuclear stellar cluster \citep{2004ApJ...613L.109P}, but is not enough to resolve
its loss cone dynamics accurately. Therefore, we do not attempt a detailed modelling of tidal 
disruption processes and set the capture radius to be equal for all particles.  
\vspace{0cm}
\begin{table}[h,t]
\vspace{0cm}
\centering{
\begin{tabular}{l||l|l|l|l|l|l}

 Runs  &      $\ \gamma$ &  $\mbul/M_{g}$  & \ \ \ \ \ $f$             &\ $\Delta$  & \ $r_h$    &  $\ln \Lambda$   \\
\toprule
\ \ $6$  &  \ \ $1$        & \ \ $0.05$	   &  $2.5\times 10^{-3}$   &    $0.08$       & $0.46$  &      $8.3$     \\
\ \ $6$  &  \ \ $1$        & \ \ $0.05$	   &  \ $5.\times 10^{-3}$  &    $0.15$       & $0.46$  &      $8.3$        \\
\ \ $6$  &  \ \ $1$        & \ \ $0.05$	   &  $7.5\times 10^{-3}$   &    $0.23$       & $0.46$  &      $8.3$       \\
\ \ $6$  &  \ \ $1$        & \ \ $0.05$	   &  \ \ \ $0.01$	               &    $0.31$       & $0.46$  &      $8.3$         \\
\ \ $4$  &  \ \ $1$        & \ \ $0.05$	   &  \ \ $0.429$	               &    $13.2$       & $0.46$  &      $8.3$          \\
\ \ $2$  &         $1/2$	   & \ \ $0.01$	   &  $2.5\times 10^{-3}$    &    $0.08$       & $0.26$  &      $7.2$  \\
\ \ $2$  &         $1/2$	   & \ \ $0.01$	   &\   $5.\times 10^{-3}$   &    $0.15$       & $0.26$  &      $7.2$    \\
\ \ $2$  &         $1/2$	   & \ \ $0.01$	   &  $7.5\times 10^{-3}$    &    $0.23$       & $0.26$  &      $7.2$    \\
\ \ $2$  &         $1/2$	   & \ \ $0.01$	   &  \ \ \ $0.01$	               &    $0.31$        & $0.26$  &      $7.2$      \\

\end{tabular}}
\vspace{0.cm}
\caption{$N$-body integrations. $1^{\rm{st}}$ column: number of runs;  $2^{\rm{nd}}$column:  slope of the Dehnen's model inner cusp at $t=0$;
 $3^{\rm{rd}}$: ratio of BH mass to total cluster mass in stars; $4^{\rm{th}}$: $f = N_H/N$ fraction of heavy mass particles; $5^{\rm{th}}$: 
Alexander \& Hopman parameter; $6^{\rm{th}}$: influence radius $r_h$; $7^{\rm{th}}$: Coulomb logarithm at $r_h$. The total number of
particles is $N=1.24 \times 10^5$ in all runs; the mass ratio between heavy and light components is $R=10$ for all runs.   The tidal 
capture radius $r_{cap}=10^{-7}$ in all runs. We use units $G=M_{nuc}=a =1$, where $M_{nuc}$ is the total mass of the nuclear cluster and $a$ is the Dehnen model's 
scale length.}
\label{tb1}
\end{table}
The MBH dominates the dynamics inside its influence radius $r_h$ defined to be the radius which
encloses twice of its mass at $t=0$. The stellar distribution evolves and reaches its asymptotic
quasi-steady state over relaxation time scales \citep{1987degc.book.....S}:
\begin{equation}
T_{rlx}(r_h) \sim 0.34 \frac{\sigma_h^3}{G^2 \rho_h m_* \ln \Lambda },
\end{equation}
where $\sigma_h$ and $\rho_h$ are, respectively, the $1D$ velocity dispersion and spatial density 
evaluated at $r_h$. Following \cite{2004ApJ...613L.109P}, we define the Coulomb logarithm 
$\ln \Lambda = \ln(r_h \sigma_h^2/2 G m_*)$.

A realistic mass population of stars with a continuous range of stellar masses can be approximately
represented by two (well-separated) mass scales: one in the range ${\cal O}(1 M_\odot)$ corresponding
to low mass main-sequence stars, white dwarfs (WDs) and neutron stars (NSs); another with 
${\cal O}(10 M_\odot)$ representing stellar black holes (SBHs). The relative abundance of objects in
these mass ranges is overwhelmingly dominated by the lighter stars --- typical number ractions of
SBHs being of ${\cal O}(10^{-3})$ \citep{2005PhR...419...65A}. 
\begin{figure}
\epsscale{1.0}
\plotone{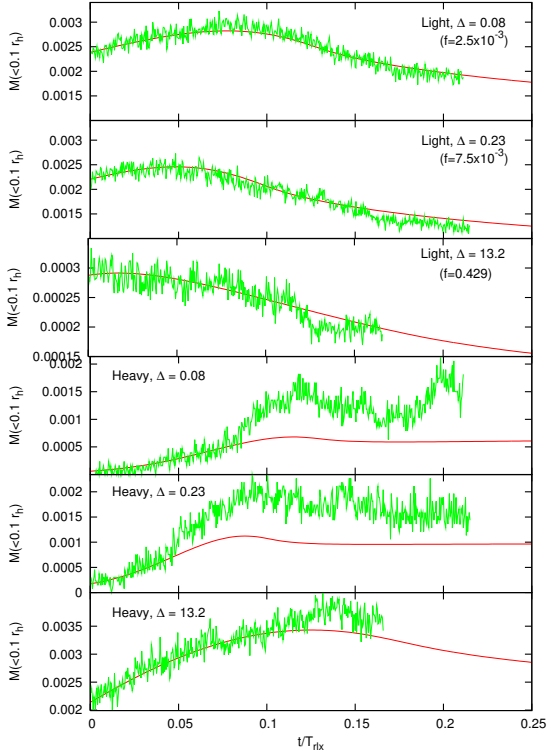}
\caption{Evolution of stellar mass within $0.1 \ r_h$ from the MBH, for light ($3$ upper panels) and
heavy ($3$ lower panels) components. Noisy curves are from $N$-body integrations; smooth curves are
from the Fokker-Planck evolution.}
\label{fig-enc-mass-vs-time}
\end{figure}
The initial stellar model is built from a Dehnen model of a spherical galaxy \citep{1993MNRAS.265..250D} 
to which a massive particle is added at the center at rest \citep{1994AJ....107..634T}. The positions and
velocities are Monte-Carlo realizations that accurately reproduce the spatial $\rho(r)$ and phase space
$f(E)$ densities with stars of the same mass. In order to generate a two-component model, we
proceed as follows: (1) we specify the mass ratio $R=m_H/m_L$ between heavy and light stars,  and 
respective number fractions $f_H=N_H/N$ and $f_L=1-f_H$, through which the AH09 $\Delta$
parameter is fixed; (2) we assign the mass $m_H$ or $m_L$ randomly to each star according to the statistical 
weights $f_H$ and $f_L$, respectively. The resulting model is almost in dynamical equilibrium; deviations
of virial ratio from unity are $\lesssim 1-2 \%$. On a dynamical time scale, phase mixing occurs 
and the virial ratio converges to unity to within a fraction of a percent. Following this prescription, the 
two-component models start without any mass segregation, as would be expected from a violently relaxed
system. Dehnen model's density has $\rho(r) \propto r^{-\gamma}$ at the center and the corresponding distribution 
function $f(E)$ is isotropic. Table 1 gives the list of runs and adopted parameters. 

\section{Fokker-Planck Models for several stellar masses}
\begin{figure}
\epsscale{1.0}
\plotone{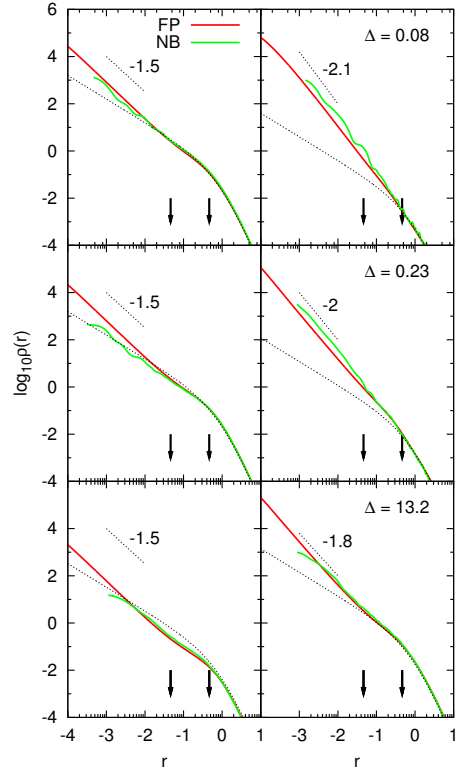}
\caption{The mass density profiles around the MBHs for both components at the end of the integrations.
$N$-body and Fokker-Planck curves are superimposed for comparison. Left panels show $\rho_L(r)$ for
the light component; right panels show $\rho_H(r)$ for heavy component.  The arrows signal the location 
of $0.1r_h$ and $r_h$ radii. These plots highlight the asymptotic
solution of both methods is in good agreement and $\rho_H \sim r^{-\gamma_H}$, where $\gamma_H$ decreases
from values $\gtrsim 2$ down to $\sim 7/4$ while moving from the strong to the weak branch of the solution.}
\label{nb-dens}
\end{figure}
We also study the evolution of the NSC with a multi-mass Fokker-Planck formalism and compare
results with the $N$-body integrations. The time-dependent, orbit-averaged, isotropic, Fokker-Planck equation
in energy space is defined, for each component \citep{1987degc.book.....S,1990ApJ...351..121C}, by

\begin{equation}
\label{fp-eqn1}
p(E) \pderiv(f_i/t) =  - \pderiv(F_{E,i}/E),\,F_{E,i}  =  -D_{EE,i} \pderiv(f_i/E) - D_E f_i,  \\
\end{equation}

\begin{eqnarray}
\label{fp-eqn2}
D_{EE,i} & = & 4\pi^2 G^2 m_*^2 \mu_i^2 \ln \Lambda  \nonumber \\
& & \times \sum_j^{N_c} \left[ \left(\frac{\mu_j}{\mu_i} \right)^2 q(E) \int_{-\infty}^E dE' f_j(E')  \right. \nonumber \\
& & \left.  + \int_E^{+\infty} dE' q(E') f_j(E') \right],  \\
D_{E,i} & = & - \sum_j^{N_c} \left( \frac{\mu_j}{\mu_i} \right) \int_E^{+\infty} dE' p(E') f_j(E').
\end{eqnarray}
In this equation, $i,j$ run from $1$ to $N_c$ (the number of mass components), $\mu_i=m_i/m_*$ where
$m_*=1/N$ is a reference mass. $p(E) = 4 \int_0^{r_{max}(E)} dr \ r^2 \sqrt{2(E-\Phi(r))} = -\partial q/\partial E$ is
the phase-space accessible to each (bound) star of specific energy $E=-v^2/2+\Phi(r)>0$ \citep{1987degc.book.....S}, 
and the total gravitational potential $\Phi(r)$ is the sum of the contribution from the nuclear cluster plus the hole.
During our simulations, the stellar distribution and its resulting gravitational potential change substantially inside 
$r_h$ only---the region over which the MBH's potential is dominant---so we keep the contribution from the stars to 
total $\Phi(r)$ fixed throughout. This system of FP equations treats self-consistently both dynamical friction and 
two-body scattering  between all components, without any further approximations other than those inherent to the 
FP formalism. In contrast with \cite{2009arXiv0909.1318M}, our treatment is not limited to early evolution where 
the heavy component is just a small 
perturbation on the (time evolving and dominant in number) light component. As a result, we can follow both weak
and strong branches of the solution throughout without restriction. We solve the FP equations~(\ref{fp-eqn1}, \ref{fp-eqn2}) using 
the \cite{1970JCoPh...6....1C} integration scheme.

\section{Results}
\begin{figure}
\epsscale{1.0}
\plotone{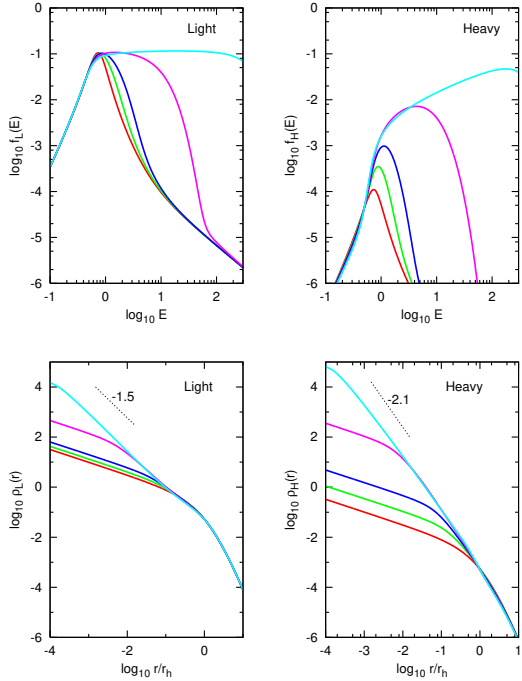}
\caption{Evolution of the phase space density (upper panels) and spatial density (lower panels)
for both components. The models starts with $\gamma=1/2$, purported to represent a cored nuclei;
the fraction of stellar black holes is $f=10^{-3}$ and their mass is $10$ times larger than that of the
light stars. Both densities increase monotonically with time and are shown at $t/T_{rlx}=0,0.05,0.1,0.2,0.25.$}
\label{fp-dens}
\end{figure}
The density of stars around the MBH increases as the cusp grows for both light and heavy 
components until it reaches a quasi-steady state; afterwards, the lights start to slowly expand. 
This can be seen in Figure~\ref{fig-enc-mass-vs-time}, where the mass
inside a sphere of $0.1 r_h$, centered on the MBH, is depicted as a function of time. This 
distance is  measured with respect to the {\it instantaneous} position of the MBH.
Both curves from FP and NB are shown together for three different runs with $\Delta=0.08,
0.23$ corresponding to strong segregation and $\Delta=13.2$ to weak segregation branches.
The time scaling between NB and FP is related through $T_{rlx}^{FP} = \ln \Lambda / N \ 
 T_{rlx}^{NB}$, and no further adjustements were made. In the three cases shown (as in all others 
cases tested but not shown), the agreement between both methods is very good, although there 
is a noticeable tendency for the heavy particles in the NB runs to segregate more strongly in the 
central cusp---this is especially the case in the strong branch. Figure~\ref{fig-enc-mass-vs-time} 
also suggests that a quasi-steady state (and maximum central concentration) have been reached by 
the end of the runs corresponding to $t \sim (0.1-0.2) T_{rlx}(r_h)$. We stress
that mass segregation, whether in the weak or strong branch, speeds up cusp growth
by factors ranging from $4$ to $10$ in comparison with the single-mass case \citep{2004ApJ...613L.109P}.

Figure~\ref{nb-dens} displays the spatial density profiles $\rho_L(r)$ and $\rho_H(r)$ at late times,
$t \sim 0.2 T_{rlx}(r_h)$. The agreement between both methods is again quite good although there is the tendency, 
in the strong branch, for NB's asymptotic slope $\gamma_L$ to be slightly smaller than in FP---for which
$\gamma_{L,min}=1.5$. The slopes of the inner  density profiles of the heavy component decrease as the solution 
evolves from the strong to the weak branch when $\Delta$ is increased, as expected. In the limit of $\Delta >> 1$, 
$\gamma_H$ tends to evolve to a quasi-steady state close to the $7/4$ solution, while for 
$\Delta << 1$, $\gamma_H \gtrsim 2$. The asymptotic inner density slopes, in both solution branches, of the 
light component extend out to $\sim 0.1 r_h$; in contrast, the heavy component shows a different behavior 
depending on the solution branch: on the weak branch, $\gamma_H$'s asymptotic slope also extends only up to 
$\sim 0.1 r_h$,  while on the strong branch it extends virtually all the way to $r_h$.  In the strong branch, the 
density of the heavy component exceeds that of the light for $r \lesssim 0.01 r_h$ (and will therefore dominate 
the interaction events with the MBH); in the strong branch, $\rho_H > \rho_L$ throughout.
\begin{figure}
\epsscale{0.75}
\rotatebox{-90}{
\plotone{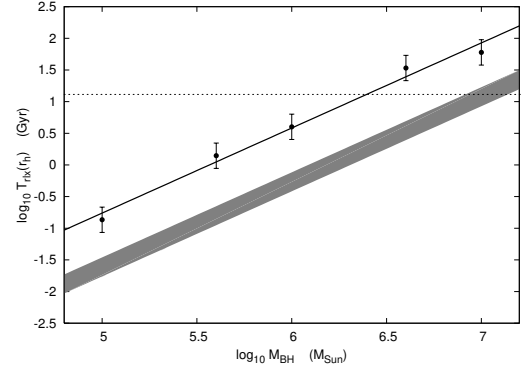}
}
\caption{The points represent the relaxation time at the influence radius $r_h$ for single mass, cored models 
of NSCs as a function of MBH mass. The shaded area covers the region $[0.1T_{rlx},0.2 T_{rlx}]$, which is time for
cusp re-growth computed with the Fokker-Planck equation. Its finite width results from the distribution 
of parameters: the number fraction of stellar BHs $f \in [10^{-3},10^{-2}]$ and mass ratio $R=10, 15, 20$. The 
horizontal curve (with points) corresponds to a Hubble time $13$ Gyr. NSCs with MBH masses below that of 
$SgrA^*$ can re-grow their stellar cusps after $\lesssim 6$ Gyr.}
\label{trlx}
\end{figure}

Although there are some differences in quantitative detail, these NB results broadly confirm the FP calculations 
and validate its inherent assumptions---at least in what concerns the description of the {\it bulk} properties of 
stellar distributions. A more detailed study and description of the stellar dynamics around a MBH under different 
initial conditions
and different models (allowing for more than the two components reported here), stellar ejections (which {\it do} 
occur in the NB runs) and captures, and further comparisons between NB and FP methods,  are outside the scope 
of this Letter and are the subject of another work in preparation.

\section{Implication for galactic nuclei and sources of GWs}
The analysis of the number counts of spectroscopically identified, old stars in the sub-parsec region of
our own Milky Way \citep{2009A&A...499..483B,2009ApJ...703.1323D}---believed to be complete down to 
magnitude $K=15.5$---reveals a deficit of old stars with respect to the high number a strongly 
segregated cusp would entail. Although the slope of the density profile is still weakly constrained, the best
fits from number counts data seem to exclude with certainty slopes $\gamma > 1$ \citep{2009A26A...502...91S},
and there could be a core with a stellar density decreasing towards  the center, $\gamma < 0$,  although such a 
fit is only marginally better than one with $\gamma \sim 1/2$.

Although we deem to be too early to conclude for the inexistence of a segregated cusp around $SgrA^*$, 
since the detectable stars (essentially giants) are still a small fraction of the stellar population as a whole, we next
compute the time necessary for cusp growth if at some point a central core is carved 
in the stellar distribution. Having validated the FP approach and its results we study equations~(\ref{fp-eqn1}, 
\ref{fp-eqn2}), which are orders of magnitude faster to solve than NB integrations. 

We choose as initial condition a model with $\gamma=1/2$ (which is the minimum slope that allows an 
isotropic solution around a MBH), since the isotropization time---the time necessary for the establishment of this 
shallow cusp starting from a hole in the spatial distribution---is $\ll T_{rlx}(r_h)$ \citep{2009arXiv0909.1318M}, and
we are interested in the evolution over $O$($T_{rlx}$) time scales. Figure~\ref{fp-dens} shows the evolving phase-space 
$f(E)$ and spatial $\rho(r)$ densities for both components ($R=10$ and $f=0.001$ constitute our fiducial case). It can 
be seen that, by $t \sim 0.25 \ T_{rlx}(r_h)$,  cusps with $\gamma_L\sim1.5$ and $\gamma_H \sim 2$ 
(or $p_L\sim 0.05$ and $p_H\sim 0.5$ in phase space) are fully developed; a little earlier, at $t \sim 0.2 \ T_{rlx}(r_h)$, 
the density cusp $\rho_H(r)$ is already fully developed down to $r \sim 0.01 r_h$ ($\sim 0.02$ pc if scaled to the Milky 
Way nucleus). If there was some event carving a hole in the stellar 
distribution around $SgrA^*$ more than $6$ Gyr ago, then there was enough time for a very steep cusp of stellar BHs 
to have re-grown.

The number fraction $f$ of SBHs is sensitive to the initial mass function (IMF) of high mass stars. There are indications 
the IMF in galactic nuclei is top-heavy \citep{2007ApJ...669.1024M} so we adopt a range of values $f \in [10^{-3},10^{-2}]$;
the mass distribution of SBHs is also weakly constrained so we follow \cite{2009MNRAS.395.2127O} in 
considering several mass ratios $R=10, 15$ and $20$. Figure~\ref{trlx} shows the relaxation times at
the influence radius for nuclei with MBH masses in the range of interest for LISA; the 
straight line is a linear fit to the points. The shaded region corresponds to the range 
$[0.1 \ T_{rlx}(r_h), 0.2 \ T_{rlx}(r_h)]$ and represents the time stellar cusps take  to grow starting from an 
isotropic core. The shaded region's width results from the distribution of values for $R$ and $f$.
In the ranges we adopted, increasing $R$ or $f$ both have the effect of decreasing the time for cusp growth. At early 
times, SBHs essentially evolve under dynamical friction with characteristic
time scale $T_{df} \sim T_{rlx}/R$; increasing $f$ leads to an increased rate of self-scattering between SBHs 
at late times.  

\section{Summary and discussion}
Our results show that strong mass segregation is a {\it robust} outcome from the
growth of stellar cusps around MBHs. We have used $N$-body integrations with
two masses---light and heavy components representing main sequence stars and
stellar BHs respectively---, and compared the results with those obtained with the
FP formalism. The broad agreement between both methods validates the FP
description of  the {\it bulk} properties of time-evolving stellar distribution around a 
MBH---and its underlying assumptions. The differences of quantitative detail are
the subject of another work in preparation. 

Using the FP equation to study cusp growth under a variety of initial conditions
purported to represent cored nuclei, we have shown that the time scales associated
with cusp re-growth are clearly shorter than a Hubble time for nuclei with MBHs
in the mass range $\mbul \lesssim 5 \times 10^6 M_\odot$---even though the relaxation
time, as estimated for a single mass stellar distribution, exceeds a Hubble time in the
upper part of this mass range. Therefore, our work 
strongly suggests that quasi-steady---strongly segregated--- stellar cusps may be 
common around MBHs with masses in this range. 

EMRIs of compact remnants will be detectable by LISA precisely for MBHs in this
mass range \citep{2006PhRvD..74b3001D,2007CQGra..24..113A,2007PhRvD..75b4005B}. Estimates
for event and detection rates by LISA costumarily assume that the stellar cusps are in
steady state \citep{2006ApJ...645.1152H,2006ApJ...645L.133H}. But recent observations
reveal a dearth of giants inside $1$ pc from $SgrA^*$ and raise the possibility that cored
nuclei are common---this scenario has been thoroughly explored by \cite{2009arXiv0909.1318M}.

Our results strongly suggest that stellar cusps can re-grow in less than a Hubble time. The  
existence of cored nuclei still remains plausible though---especially for nuclei with MBHs in
the upper part of the mass range---, since time scales are still quite long ({\it e.g.} $6$ Gyr in
Milky Way type nuclei). However, since EMRI rates scale as $\mbul^{-\alpha}$, 
$\alpha \in [\frac{1}{4},1]$,  and re-growth times are $\lesssim 1$ Gyr for 
$\mbul \lesssim 1.2 \times 10^6 M_\odot$, we still expect that a substantial fraction of EMRI 
events will originate from segregated stellar cusps.
Finally, indirect observations alone will reveal whether there is a ``hidden'' cusp of old stars 
and their dark remnants around $SgrA^*$ \citep{2005ApJ...622..878W,2009ApJ...703.1743P}.

\ \ \ \ \ 

\acknowledgements 

MP and PAS acknowledge support by DLR (Deutsches Zentrum f\"ur Luft- und
Raumfahrt). The simulations have been carried out on the dedicated high-performance
GRAPE-6A clusters at the Astronomisches Rechen-Institut in Heidelberg; \footnote{GRACE: 
see http://www.ari.uni-heidelberg.de/grace} some of the simulations were done at the 
{\sc Tuffstein} cluster of the AEI.

\end{document}